\documentclass[aps,prapplied,twocolumn,superscriptaddress,nofootinbib,longbibliography,floatfix]{revtex4-2}
\usepackage[utf8]{inputenc}
\usepackage[T1]{fontenc}
\usepackage{amsmath,amssymb,graphicx,booktabs,xcolor,microtype,url}
\usepackage[colorlinks,allcolors=blue]{hyperref}
\hypersetup{pdftitle={Native higher-order optimization with entropy quantum computing},pdfauthor={Narendra N. Hegade, Pranav Chandarana, Sebastian Wagner, Bahman Qawami, Wesley Dyk}}
\mathchardef\UrlBigBreakPenalty=100
\mathchardef\UrlBreakPenalty=5000

\graphicspath{{figs/}}
\newcommand{\Dirac}{Dirac-3S}
\newcommand{\Series}{Dirac-3}
\newcommand{\Ek}[1]{\mathbb{E}\!\left[E^{(#1)}_{\min}\right]}
\newcommand{\MVSKREACH}{27}
\newcommand{\pr}[2]{\parbox[t]{#1}{\raggedright #2}}

\begin{document}

\title{Native higher-order optimization with entropy quantum computing}

\author{Narendra N. Hegade}
\email{narendra.hegade@kipu-quantum.com}
\affiliation{Kipu Quantum GmbH, Greifswalderstrasse 212, 10405 Berlin, Germany}
\affiliation{IDAL, Electronic Engineering Department, ETSE-UV, University of Valencia, Avgda. Universitat s/n, 46100 Burjassot, Valencia, Spain}
\author{Pranav Chandarana}
\affiliation{Kipu Quantum GmbH, Greifswalderstrasse 212, 10405 Berlin, Germany}
\affiliation{Department of Physical Chemistry, University of the Basque Country UPV/EHU, 
48080 Bilbao, Spain}
\author{Sebastian Wagner}
\affiliation{Kipu Quantum GmbH, Greifswalderstrasse 212, 10405 Berlin, Germany}
\author{\\ Bahman Qawami}
\affiliation{Quantum Computing Inc., 5 Marine View Plaza, Hoboken, New Jersey 07030, USA}
\author{Wesley Dyk}
\email{wdyk@quantumcomputinginc.com}
\affiliation{Quantum Computing Inc., 5 Marine View Plaza, Hoboken, New Jersey 07030, USA}

\date{\today}

\begin{abstract}
Higher-order interactions are common in applied optimization: parity checks, radar-code sidelobes, portfolio skewness and kurtosis, and marketing-channel synergies all yield cubic or quartic objectives. Quantum annealers and most Ising machines accept only quadratic objectives, and gate-based quantum processors build every higher-order term from two-qubit gates. Entropy quantum computing, realized in the hybrid photonic--electronic \Dirac{} system, accepts polynomials of degree up to five with all-to-all coupling over binary, integer or simplex-constrained continuous variables. No user-supplied quadratization or circuit compilation is needed. We formulate four industrial use cases as six cubic and quartic instances, run them on \Dirac{} with 30 samples each, and score every sample against a certified optimum or a classical best-known value. We also estimate their cost on quantum annealers, superconducting grids and trapped-ion registers under standard mappings. For the two densest instances, a portfolio with about 80{,}000 terms and a budget allocation with 360{,}000 terms, the quadratized problems need embeddings far beyond current annealers, over $10^5$ physical qubits for the portfolio alone, and one phase-separation layer costs over $10^6$ two-qubit gates before optimization. \Dirac{} ran all six natively and returned feasible, distinct solutions. All samples of the binary instances are single-flip local minima. The best portfolio sample lies within 0.32\% of the best-known value and the best budget sample 15\% above it, but no sample reached its reference. Classical solvers reach the references within minutes at these sizes, and random starts followed by a local polish reach comparable or better values, so the advantage established here is one of expressibility and mapping cost, not of solution quality. Whether it extends to time to solution at larger sizes remains open.\looseness=-1
\end{abstract}

\maketitle

\section{Introduction}
\label{sec:intro}

Quantum annealers and most coherent Ising machines, digital annealers and simulated-bifurcation machines minimize an Ising or quadratic unconstrained binary optimization (QUBO) objective \cite{lucas2014ising,mohseni2022ising}, and the native entangling gates of most digital quantum processors act on two qubits at a time. Real objectives, however, are often not pairwise. A parity check on three bits is already a three-body term \cite{kowalsky2022xorsat}. The sidelobe energy of a binary radar code is a sum of squared correlations, so it is quartic \cite{bernasconi1987labs,packebusch2016labs}. A portfolio that accounts for skew and tail risk through co-skewness and co-kurtosis is cubic and quartic in the holdings \cite{jondeau2006higher,harvey2010higher}, and a budget split across channels that reinforce or cannibalize one another three at a time is cubic in the shares. In all four cases the objective is a polynomial of degree three or four in binary, integer or continuous variables. Such higher-order problems, called higher-order unconstrained binary optimization (HUBO) when the variables are binary, have become a target of gate-based quantum optimization through the quantum approximate optimization algorithm (QAOA) \cite{shaydulin2024labs} and digitized counterdiabatic algorithms \cite{cadavid2025bfdcqo,romero2025bfdcqo_hubo,chandarana2025runtime,cadavid2025qemts}. Higher-order coupling is also being explored in Ising machines, from optical emulators with tailored four-body interactions \cite{kumar2020fourbody} to coupled-oscillator, neuromorphic and probabilistic-bit machines \cite{bybee2023higher,ahsan2026neuromorphic,nikhar2024alltoall}, but the widely deployed quantum annealers remain quadratic.

Bridging the gap between such a model and a pairwise machine is expensive. A quadratic annealer needs the higher-order terms quadratized. Products of variables become auxiliary variables held in place by penalty terms \cite{rosenberg1975reduction,boros2002pseudo,dattani2019quadratization}, which increases the number of variables, widens the coefficient range and adds spurious local minima to the landscape. The dense QUBO that results must then be embedded into a hardware graph whose vertices have 15 or 20 neighbors \cite{boothby2020pegasus,boothby2021zephyr}, with chains of physical qubits standing in for every logical one \cite{choi2008minor}. Both overheads have been quantified in general \cite{valiante2021locality,konz2021embedding}; Sec.~\ref{sec:resources} evaluates them for our instances. A gate-based processor needs no auxiliaries, but every $w$-body term becomes a Pauli-$Z$ string rotation that costs $2(w-1)$ two-qubit gates, and on a square-grid or heavy-hex lattice the qubits involved must first be brought together with SWAP gates. Trapped-ion registers avoid the routing, but hold about a hundred fully connected qubits, and their two-qubit gates take of order a hundred microseconds. Even on classical computers, a simulated-annealing solver that keeps the polynomial intact has been reported to outperform its QUBO counterpart \cite{ikeuchi2025pubo}, a sign that the reduction itself, and not only the device, is part of the problem.

Entropy quantum computing, the paradigm behind the \Series{} series of hybrid photonic--electronic optimizers \cite{nguyen2024entropy}, avoids the reduction altogether. These devices accept polynomials of degree up to five natively, with all-to-all coupling, over binary, integer or simplex-constrained continuous variables, so a cubic or quartic objective is submitted without auxiliary variables or penalty terms. Published results on these devices so far include the first demonstrations \cite{nguyen2024entropy}, a protein-design application \cite{emami2026protein}, a quartic matrix-factorization formulation over continuous and integer variables compared with a classical library routine \cite{borle2025nmf}, a Motzkin--Straus formulation of clique problems compared with classical continuous optimizers \cite{wang2026motzkin}, and an independent critical assessment of the first demonstrations \cite{moosavian2026critical}. What is missing is a set of higher-order use cases spanning binary, integer and continuous variables in which every returned solution is scored against a certified optimum or an independently established best-known value, and an account of what the same instances would cost on quantum annealers and gate-based processors. Here we present such use cases and the results obtained for them on \Dirac{}, the current generation of the series. The device runs every instance in its native form and returns feasible solutions, but no sample reaches its reference: the advantage we find is one of expressibility and mapping cost; in solution quality we find none.

The paper makes three contributions. Section~\ref{sec:usecases} gives explicit higher-order formulations of the four use cases, with instances whose reference values are certified or independently established. After a description of the device in Sec.~\ref{sec:eqc}, Sec.~\ref{sec:resources} quantifies what the same instances would cost on a quadratic annealer and on superconducting and trapped-ion processors. Sections~\ref{sec:results} and \ref{sec:discussion} report the device results, in which every sample is scored independently and compared with matched classical controls, and draw the practical consequences. Protocol, scoring and reference computations are in Appendix~\ref{app:methods}. Classical computation supplies the reference values, the cheap polish applied to the device samples, and matched-start controls for that polish; no time-to-solution comparison is made.

\section{Use cases and their higher-order formulations}
\label{sec:usecases}

The four use cases sketched in the introduction are worked out here as explicit polynomials and turned into the six instances used throughout the paper. They come from industrial practice in coding and cryptanalysis, radar and communications, asset management and marketing; the instances are test problems of these types, two of them built on synthetic data. Table~\ref{tab:usecases} summarizes the use cases and the origin of their cubic or quartic terms, Table~\ref{tab:instances} lists the resulting instances, and Appendix~\ref{app:instances} gives the construction details.

\begin{table*}[t]
\caption{The four use cases, their decision variables, and the origin of the higher-order terms. Every objective is minimized. The last column gives the constraint and the device mode used on \Dirac{}.}
\label{tab:usecases}
\begin{ruledtabular}
\begin{tabular}{p{3.6cm}p{2.5cm}p{1.1cm}p{5.5cm}p{3.5cm}}
Use case & Variables & Degree & Origin of the degree-3/4 terms & Constraint / mode \\
\hline
\pr{3.5cm}{Parity-constrained satisfiability (3-XORSAT): decoding, cryptanalysis} & \pr{2.4cm}{$N$ bits $x_i\in\{0,1\}$} & \pr{1.0cm}{3} & \pr{5.4cm}{each check $x_a\oplus x_b\oplus x_c$ contributes $1-\sigma_m s_as_bs_c$ in $s_i=1-2x_i$, with $\sigma_m=\pm1$ the required parity} & \pr{3.4cm}{none (binary, integer mode)} \\[3pt]
\pr{3.5cm}{Low-autocorrelation sequence design (radar, communications)} & \pr{2.4cm}{$N$ signs $s_i=\pm1$} & \pr{1.0cm}{4} & \pr{5.4cm}{$E=\sum_k C_k^2$ with bilinear correlations $C_k=\sum_i s_is_{i+k}$} & \pr{3.4cm}{none (binary, integer mode)} \\[3pt]
\pr{3.5cm}{Higher-moment portfolio with round lots (asset management)} & \pr{2.4cm}{$N$ lots $x_i\in\{0,\dots,3\}$} & \pr{1.0cm}{4} & \pr{5.4cm}{co-skewness $\sum S_{ijk}x_ix_jx_k$ and co-kurtosis $\sum K_{ijkl}x_ix_jx_kx_l$ of returns} & \pr{3.4cm}{$0\le x_i\le 3$ (integer mode, 4 levels)} \\[3pt]
\pr{3.5cm}{Budget allocation with channel synergies (marketing mix)} & \pr{2.4cm}{$N$ shares $v_i\ge0$} & \pr{1.0cm}{3} & \pr{5.4cm}{three-channel synergies $\sum_{i<j<k}g_{ijk}v_iv_jv_k$ with saturation $b_iv_i^2$} & \pr{3.4cm}{$\sum_i v_i=1$ on the device (continuous simplex mode)} \\
\end{tabular}
\end{ruledtabular}
\end{table*}

\begin{table*}[t]
\caption{The six instances. Terms are counted by degree (linear/quadratic/cubic/quartic). The coefficient dynamic range is the ratio of the largest to the smallest absolute coefficient of the submitted polynomial. References are certified optima (3-XORSAT, LABS) or best-known values (MVSK, budget), whose status is described in the text; $E^*$ and $F^*$ are the optimal LABS energy and merit factor; SA stands for simulated annealing.\looseness=-1}
\label{tab:instances}
\begin{ruledtabular}
\begin{tabular}{llcrlrl}
Instance & Variables & Degree & Terms & Terms by degree & Dyn.\ range & Reference \\
\hline
3-XORSAT $N{=}100$ & 100 binary & 3 & 493 & 100/293/100 & 4 & 0 unsat.\ (planted) \\
LABS $N{=}24$ & 24 binary & 4 & 3{,}246 & 24/276/2{,}000/946 & 45 & $E^*{=}36$ ($F^*{=}8.00$) \\
LABS $N{=}28$ & 28 binary & 4 & 5{,}197 & 28/378/3{,}244/1{,}547 & 63 & $E^*{=}50$ ($F^*{=}7.84$) \\
LABS $N{=}34$ & 34 binary & 4 & 9{,}387 & 34/561/5{,}936/2{,}856 & 96 & $E^*{=}65$ ($F^*{=}8.89$) \\
MVSK $N{=}35$ & 35 integer $\{0,\dots,3\}$ & 4 & 79{,}164 & 35/617/6{,}607/71{,}905 & 997 & $-6{,}867{,}311$ (SA) \\
Budget $N{=}130$ & 130 continuous, $\sum v_i{=}1$ & 3 & 356{,}570 & 130/130/356{,}310 & 1{,}000 & $-4.3495$ (seeded SLSQP) \\
\end{tabular}
\end{ruledtabular}
\end{table*}

\subsection{Parity-constrained satisfiability (3-XORSAT)}
A parity check on three bits, $x_a \oplus x_b \oplus x_c$ fixed to 0 or 1, is the elementary constraint of binary linear codes, of syndrome decoding and of the algebraic systems that arise in cryptanalysis. In spin variables $s_i = 1 - 2x_i$, check $m$ holds if and only if $\sigma_m s_a s_b s_c = 1$, where $\sigma_m = +1$ if the check requires even parity and $-1$ if it requires odd parity. Each check contributes $1 - \sigma_m s_a s_b s_c$, zero when satisfied and two when violated, so the sum over the $M$ checks of the instance,
\begin{equation}
E(s) = \sum_{m=1}^{M} \left(1 - \sigma_m\, s_{a_m} s_{b_m} s_{c_m}\right),
\label{eq:xorsat}
\end{equation}
is a cubic polynomial with integer coefficients and one three-body term per check, and the figure of merit $E/2$ is the number of violated checks. No quadratic polynomial in the original bits reproduces three-bit parity, so a QUBO needs up to one auxiliary variable per check. Our instance has $N{=}100$ bits and $M{=}100$ checks, each variable appearing in exactly three checks, with the parities $\sigma_m$ chosen so that a planted assignment satisfies every check. The optimum, zero violated checks, is therefore certified by construction. Gaussian elimination solves the problem in polynomial time. The landscape of planted 3-regular 3-XORSAT, however, is of the golf-course type, flat except for a narrow well at the planted solution. This makes it a standard hard test problem for annealers and local heuristics, whose time to solution grows exponentially in $N$ \cite{kowalsky2022xorsat}.\looseness=-1

\subsection{Low-autocorrelation binary sequences}
A pulse-compression radar or a spread-spectrum link transmits a binary phase code $s \in \{\pm1\}^N$ and correlates the return against it. The off-peak values of the aperiodic autocorrelation $C_k = \sum_{i=1}^{N-k} s_i s_{i+k}$ appear as range sidelobes that mask weak targets or other users. The low-autocorrelation binary sequence (LABS) problem \cite{golay1977sieves,bernasconi1987labs} asks for the code that minimizes the sidelobe energy
\begin{equation}
E(s) = \sum_{k=1}^{N-1} C_k^2 = \sum_{k=1}^{N-1}\Big(\sum_{i=1}^{N-k} s_i s_{i+k}\Big)^2,
\label{eq:labs}
\end{equation}
usually quoted as the merit factor $F = N^2/2E$. Each $C_k$ is bilinear, so $E$ is a quartic polynomial with $O(N^3)$ four-body terms and all-to-all coupling. The quartic degree is intrinsic, since the figure of merit squares the correlations. Exhaustive branch-and-bound has certified the optimal energies for $N \le 66$ \cite{mertens1996exhaustive,packebusch2016labs}, which makes LABS one of very few higher-order problems with known optima at sizes that current devices can hold. It is the higher-order problem of the Quantum Optimization Benchmarking Library \cite{koch2025qoblib} and the problem on which a scaling advantage of QAOA has been argued \cite{shaydulin2024labs}. A favorable heuristic scaling has been reported for quantum-enhanced memetic tabu search \cite{cadavid2025qemts}, built on the classical memetic tabu search \cite{gallardo2009memetic,boskovic2017labs}. We use $N{=}24$, 28 and 34 (Table~\ref{tab:instances}).

\subsection{Higher-moment portfolio with round lots}
Mean--variance portfolio selection \cite{markowitz1952portfolio} describes returns by their first two moments. Real returns are asymmetric and fat-tailed, investors pay for positive skewness and avoid kurtosis, and the natural extension is a mean--variance--skewness--kurtosis (MVSK) objective \cite{jondeau2006higher,harvey2010higher}. Because holdings are traded in round lots, the decision variables are small integers. Our instance is a synthetic allocation model of this type: it has a return history, a reward vector drawn independently of it and a lot box, but no prices, capital constraint or financing rule. For $N{=}35$ assets held in $x_i \in \{0,1,2,3\}$ lots we minimize
\begin{align}
E(x) = {} & -\lambda_1 \mu^{\!\top} x + \lambda_2 \sum_{ij} \Sigma_{ij}\, x_i x_j \nonumber \\
 & - \lambda_3 \sum_{ijk} S_{ijk}\, x_i x_j x_k + \lambda_4 \sum_{ijkl} K_{ijkl}\, x_i x_j x_k x_l ,
\label{eq:mvsk}
\end{align}
with $\lambda = (8, 1, 0.3, 0.1)$, where $\mu$ is a reward vector and $\Sigma$, $S$ and $K$ are the sample central co-moment tensors of a synthetic $T{=}600$-observation return history with fat-tailed and asymmetric shocks (Appendix~\ref{app:instances}). The co-skewness and co-kurtosis terms couple three and four assets at a time. They are part of the model itself, and replacing them by a quadratic surrogate changes the objective being optimized. Without a capital constraint, the quartic term penalizes large holdings, so the best-known allocation is nontrivial and uses all four lot levels.\looseness=-1

The full unrounded objective is a convex function of $x$ (Appendix~\ref{app:instances}), so the minimum of its continuous relaxation over the box $[0,3]^{35}$ can be computed with a certificate and bounds the integer optimum from below. The difficulty of that formulation lies in the integrality of the lots and, for a device, in the dense quartic coupling.

On \Dirac{} the lots map directly onto integer variables with four levels. Coefficients are scaled and rounded to integers, and terms below $10^{-3}$ of the largest are dropped to limit the dynamic range (Table~\ref{tab:instances}). Rounding and pruning do not in general preserve convexity, and no certificate exists for the submitted polynomial. Its best-known objective, $-6{,}867{,}311$, was reached by all five seeds of a problem-tuned simulated annealing. The same lot vector is the best known for the unpruned polynomial, on which an independent integer parallel-tempering code came within 0.6\% of it and found nothing lower. On the unrounded objective it lies 2.7\% above the certified relaxation bound (Appendix~\ref{app:instances}).

\subsection{Budget allocation with channel synergies}
A marketing budget, an advertising plan or a research portfolio distributes a fixed total over $N$ channels. Each channel saturates, and channels interact: a television campaign amplifies a search campaign, and two similar channels cannibalize each other. Interactions among three channels are cubic in the shares. With $v_i \ge 0$ and $\sum_i v_i = 1$ we minimize the negative response
\begin{equation}
E(v) = -\sum_i a_i v_i + \sum_i b_i v_i^2 - \sum_{i<j<k} g_{ijk}\, v_i v_j v_k ,
\label{eq:budget}
\end{equation}
with uplifts $a_i$, saturations $b_i > 0$ and signed synergies $g_{ijk}$, all drawn at random (Appendix~\ref{app:instances}); the instance is a synthetic stress test of a dense cubic objective under the simplex constraint, not a calibrated marketing model. The signed cubic term makes the problem non-convex, with many strict local minima on the simplex. The budget constraint is the native simplex constraint of the device, whereas the classical solver handles it with sequential least-squares quadratic programming (SLSQP) \cite{kraft1988slsqp,virtanen2020scipy}. Synergies below $10^{-3}$ of the largest coefficient were zeroed for the same reason as in the portfolio (Table~\ref{tab:instances}), and the SLSQP reference optimizes the identical sparsified instance. The best-known objective is $E = -4.349464$, a three-channel allocation found by seeded SLSQP. It is certified as the minimum over all allocations that fund at most three channels (Appendix~\ref{app:instances}). Allocations with four or more channels have not been searched exhaustively, so for the full instance it remains a best-known value.

\section{Entropy quantum computing and the \Dirac{} system}
\label{sec:eqc}

\begin{figure}[t]
\centering
\includegraphics[width=\columnwidth]{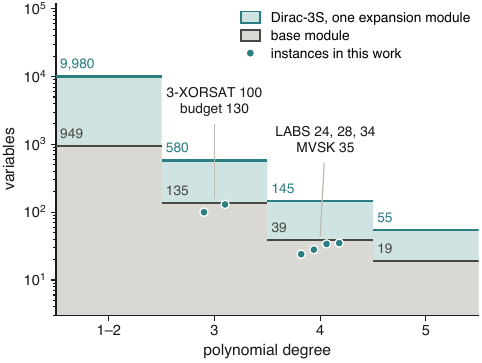}
\caption{Native problem envelope: maximum number of variables as a function of the polynomial degree of the objective, for \Dirac{} with one expansion module (outer region) and for the base module alone (inner region), with the instances of this study (dots). The instances were sized to the base-module envelope, of which the largest cubic and quartic instances fill 96\% and 90\%; they occupy less than a quarter of the \Dirac{} envelope. The envelope is the specified capacity and was not itself demonstrated in this work.}
\label{fig:envelope}
\end{figure}

The \Series{} systems are hybrid photonic--electronic optimizers \cite{nguyen2024entropy}. In continuous mode each decision variable is assigned a time bin of a weak coherent pulse train that circulates in a fiber loop; in integer mode a variable with $L$ levels is a qudit encoded in $L$ time bins \cite{qci_dirac3_guide,qci_dirac3s_guide}. The photon number in bin $i$ represents $v_i$, the constraint $\sum_i v_i = R$, with a user-set total $R$ ($R=1$ here), is realized through the total photon number of the train, and the returned occupations are normalized to $R$ \cite{qci_dirac3_guide}. On each round trip, single-photon detection with time-correlated counting measures the occupation of every bin, and a field-programmable gate array turns the counts into a bin-dependent loss for the next pass. The loss on bin $i$ has a constant part, the linear coefficient of the objective, and photon-number-dependent parts that encode the quadratic to quintic couplings, so that configurations of higher objective value are attenuated more strongly. Poisson fluctuations of the single-photon counts provide a controlled noise that lets the system leave shallow minima. Over the feedback iterations of a relaxation schedule the low-loss configuration accumulates amplitude. Reference~\cite{nguyen2024entropy} describes the process as conditioning the ground state through dissipation in an open system rather than through the adiabatic evolution of a closed one. The interactions are thus computed electronically from measured photon counts, while the optical loop supplies the parallel sampling of all bins and the shot-noise statistics of the counts. The same reference notes that this measurement-feedback implementation could in principle be simulated classically. We use the established name of the paradigm and make no claim that a nonclassical resource contributes to the results reported here.

For the user the device minimizes a polynomial of degree up to five with real coefficients $c$ and all monomials allowed,
\begin{equation}
\sum_{i} c_i v_i + \sum_{i\le j} c_{ij} v_i v_j + \sum_{i\le j\le k} c_{ijk} v_i v_j v_k + \cdots,
\label{eq:poly}
\end{equation}
either over the continuous simplex $v_i \ge 0$, $\sum_i v_i = R$, or over integers with a per-variable number of levels, two for binary variables \cite{qci_dirac3_guide}. A relaxation schedule, numbered 1 to 4, selects a preset of loss, number of feedback iterations, quantum-fluctuation coefficient and mean photon number. Higher schedules evolve more slowly and can give better solutions at the cost of longer device time \cite{qci_dirac3_guide}. Each requested sample is a separate run of the loop. The user guide puts the effective analog resolution of the coefficients at about 200:1 \cite{qci_dirac3s_guide}. The 3-XORSAT and LABS polynomials lie within this range, whereas the MVSK and budget polynomials span about 1{,}000:1 (Table~\ref{tab:instances}), so their smallest terms may not be resolved on the device. For the budget instance these terms are few and light: removing them changes the best-known objective by 1.5\% and leaves the ranking of the samples nearly intact. For the portfolio they are 14\% of the terms and matter in aggregate: removing them changes the best-known objective by 20\% and reorders the samples. How well the device resolved them cannot be read from the samples, and every sample is scored on the full submitted polynomial.

\Dirac{} consists of a base module to which expansion modules can be added to increase the problem-size capacity \cite{qci_dirac3s_guide}. The system used in this paper has a single expansion module, which raises the capacity to 9{,}980 variables for quadratic objectives and to 580, 145 and 55 variables for degrees three, four and five. The base module alone holds 949, 135, 39 and 19, and its integer mode is further limited to 949 levels summed over all variables \cite{qci_dirac3_guide,qci_dirac3s_guide} (Fig.~\ref{fig:envelope}). The instances of this study were sized for the base-module envelope and occupy 48 to 200 time bins. They were run on 28~August~2026 at relaxation schedule~3 with 30 samples per instance. Total compute time consists of preprocessing, runtime and postprocessing \cite{qci_dirac3s_guide}. The job record gives the device time as a single number without this breakdown, and the separately reported calibration time is excluded. The user guide lists an error-correction step in the postprocessing of every sample and distinguishes raw from corrected solutions \cite{qci_dirac3s_guide}, without specifying the correction, and the job record holds one solution per sample. The local-optimality results below therefore characterize the returned configurations and cannot be attributed to the photonic dynamics alone.

\section{Resource requirements on quantum annealers and gate-based processors}
\label{sec:resources}

\begin{table*}[t]
\caption{What the six instances require natively on \Dirac{} and on a quadratic annealer. Native: variables, polynomial terms and measured device time per sample (schedule 3). Annealer: binary variables after encoding (2 bits per MVSK asset; 4- or 8-bit fixed point per budget channel), auxiliary variables of the pair-cover quadratization, logical variables and couplers of the resulting QUBO, the ratio of the largest penalty weight to the largest coefficient before quadratization (not evaluated for the budget instance), and a lower bound on the physical qubits needed to embed that QUBO into a hardware graph of vertex degree 15. Current annealers provide about 4{,}400 physical qubits at vertex degree 20 and about 5{,}600 at degree 15 \cite{boothby2020pegasus,boothby2021zephyr,dwave_qpu_docs,dwave2025adv2}. Auxiliary counts are upper bounds on the minimum, and the embedding bounds refer to the quadratization used here, not to every possible encoding (Appendix~\ref{app:resources}).\looseness=-1}
\label{tab:res_annealer}
\footnotesize
\begin{ruledtabular}
\begin{tabular}{lrrrrrrrrr}
 & \multicolumn{3}{c}{Native (\Dirac{})} & \multicolumn{6}{c}{Quadratic annealer (QUBO + degree-15 hardware graph)} \\
Instance & Vars & Terms & s/sample & Binary vars & Auxiliaries & Logical vars & Couplers & Penalty ampl. & Phys.\ qubits $\ge$ \\
\hline
3-XORSAT $N{=}100$ & 100 & 493 & 2.09 & 100 & 95 & 195 & 585 & 2.2$\times$ & 195 \\
LABS $N{=}24$ & 24 & 3{,}246 & 0.27 & 24 & 259 & 283 & 3{,}740 & 2.7$\times$ & 697 \\
LABS $N{=}28$ & 28 & 5{,}197 & 0.51 & 28 & 358 & 386 & 5{,}885 & 2.3$\times$ & 1{,}068 \\
LABS $N{=}34$ & 34 & 9{,}387 & 1.00 & 34 & 536 & 570 & 10{,}425 & 1.9$\times$ & 1{,}829 \\
MVSK $N{=}35$ & 35 & 79{,}164 & 4.70 & 70 & 2{,}415 & 2{,}485 & 953{,}755 & 155.1$\times$ & 147{,}520 \\
Budget $N{=}130$ (4-bit) & 130 & 356{,}570 & 42.07 & 520 & 66{,}560 & 67{,}080 & 23{,}071{,}900 & -- & 3{,}539{,}204 \\
Budget $N{=}130$ (8-bit) & 130 & 356{,}570 & 42.07 & 1{,}040 & 266{,}240 & 267{,}280 & 183{,}503{,}480 & -- & 28{,}190{,}185 \\
\end{tabular}
\end{ruledtabular}
\end{table*}

\begin{table*}[t]
\caption{What the six instances require on gate-based processors, per layer of the phase-separation unitary $e^{-i\gamma H}$ of the cost Hamiltonian $H$ with layer angle $\gamma$ (one layer of QAOA or of a digitized counterdiabatic circuit). $H$ is a sum of Pauli-$Z$ strings, and a string of weight $w$ costs $2(w-1)$ two-qubit gates on all-to-all connectivity, the count for a trapped-ion register. On a square-grid superconducting lattice the partner qubits are first routed next to a target by SWAP gates. All counts are for independent synthesis of each string: they are construction counts, not lower bounds, and the routing is a simple, unoptimized per-string model (Appendix~\ref{app:resources}). Layer times assume serial execution at 200~$\mu$s (trapped ion) and 0.3~$\mu$s (superconducting) per two-qubit gate. The last column is the number of layers whose all-to-all gate count fits into $10^{3}$ two-qubit gates, the inverse of a representative two-qubit error of $10^{-3}$; this is a noise scale, not a success threshold.}
\label{tab:res_gate}
\footnotesize
\begin{ruledtabular}
\begin{tabular}{lrrrrrr}
 & & \multicolumn{2}{c}{Trapped ion (all-to-all)} & \multicolumn{2}{c}{Superconducting (square grid)} & \\
Instance & Qubits & 2q gates / layer & time / layer & 2q gates / layer & time / layer & Layers per $10^3$-gate noise scale \\
\hline
3-XORSAT $N{=}100$ & 100 & 400 & 0.08 s & 3{,}760 & 0.0011 s & 2.5 \\
LABS $N{=}24$ & 24 & 5{,}940 & 1 s & 25{,}542 & 0.0077 s & 0.17 \\
LABS $N{=}28$ & 28 & 9{,}646 & 2 s & 51{,}445 & 0.015 s & 0.1 \\
LABS $N{=}34$ & 34 & 17{,}680 & 4 s & 94{,}293 & 0.028 s & 0.057 \\
MVSK $N{=}35$ & 70 & $5.6\times10^{6}$ & 19 min & $4.7\times10^{7}$ & 14 s & $2\times10^{-4}$ \\
Budget $N{=}130$ (4-bit) & 520 & $9.1\times10^{7}$ & 5 h & $2.1\times10^{9}$ & 10 min & $10^{-5}$ \\
Budget $N{=}130$ (8-bit) & 1{,}040 & $7.3\times10^{8}$ & 41 h & $2.4\times10^{10}$ & 119 min & $10^{-6}$ \\
\end{tabular}
\end{ruledtabular}
\end{table*}

We estimate what the same six instances would require on three quantum platforms used for optimization: quadratic annealers, gate-based superconducting processors and trapped-ion processors. The estimates refer to specific mappings, quadratization followed by minor embedding for the annealer and circuits that implement the phase-separation unitary $e^{-i\gamma H}$ of the cost Hamiltonian $H$ for the gate-based processors; the models are stated in full in Appendix~\ref{app:resources}. Variational circuits that sample bitstrings and evaluate the polynomial classically \cite{barkoutsos2020cvar}, or compiled objective-evaluation circuits \cite{bako2025progqaoa}, are different routes and are not bounded by our numbers. Within these mappings, the embedding figures are lower bounds for the specified QUBO graphs, whereas the auxiliary counts and penalty weights are sufficient rather than minimal. The circuit counts are construction counts for independent synthesis of each Pauli string, which compilation can lower, and include no error mitigation or repetition overhead. Tables~\ref{tab:res_annealer} and \ref{tab:res_gate} collect the numbers.

\paragraph{Quadratic annealers.}
Mapping the model onto the machine incurs costs at three steps: binary encoding, quadratization and embedding. The four-level portfolio needs two bits per asset, which turns its 79{,}164 integer monomials into nearly a million multilinear bit-monomials. The continuous budget shares need a fixed-point expansion of 4 or 8 bits per channel plus a penalty $(\sum_i v_i - 1)^2$ that couples every bit to every other. Their cubic term alone expands into more than $10^7$ bit-monomials, and the fixed-point weights widen its coefficient range.

Quadratization then adds auxiliaries. Each cubic monomial needs one auxiliary variable standing for the product of two bits and each quartic monomial two, held in place by Rosenberg penalties \cite{rosenberg1975reduction}. Even with auxiliaries shared across monomials, the quadratized quartic instances have 12 to 71 times as many logical variables as the native ones, and our construction for the budget instance uses $7\times10^4$ to $3\times10^5$ auxiliaries (Table~\ref{tab:res_annealer}). We set each penalty weight just above the total coefficient weight routed through its auxiliary, a sufficient choice, and this stretches the coefficient range. The final LABS QUBOs span coefficient ranges of 370 to 550. For the portfolio the bit expansion alone widens the range from $10^3$ to $10^6$, and the penalties raise it to $4\times10^8$ (Appendix~\ref{app:resources}). An analog annealer whose coefficient resolution is of order 1\% of the largest coupler offers roughly $10^2$ distinguishable levels, a representative figure and not a sharp cutoff.

Embedding follows. The quadratized graphs are dense: the portfolio has 2{,}485 logical variables with couplers on 31\% of all pairs. On a hardware graph of vertex degree 15, a logical node of degree $\delta$ needs a chain of at least $\max\{1, \lceil(\delta-2)/13\rceil\}$ physical qubits (Appendix~\ref{app:resources}). The resulting lower bounds are 697 to 1{,}829 qubits for LABS, 147{,}520 for the portfolio and $3.5\times10^6$ for the 4-bit budget instance, against the 4{,}400 to 5{,}600 qubits of current annealers \cite{dwave_qpu_docs,dwave2025adv2}. These bounds count coupler endpoints only: they are necessary conditions, and heuristic embedders generally need more. The sparse 3-XORSAT instance is the exception. Its 195 logical variables have maximum degree 9 and embed with short chains, which is why this problem class was run on annealers before \cite{kowalsky2022xorsat}.\looseness=-1

\paragraph{Gate-based superconducting processors.}
Higher-order terms need no auxiliaries here. In spin variables the cost Hamiltonian is a sum of Pauli-$Z$ strings, and a string of weight $w$ becomes a rotation compiled into $2(w-1)$ CNOT gates and one single-qubit rotation. With all-to-all connectivity, one layer of $e^{-i\gamma H}$ contains 17{,}680 two-qubit gates for LABS $N{=}34$, $5.6\times10^6$ for the portfolio and $9\times10^7$ for the 4-bit budget instance (Table~\ref{tab:res_gate}). Placing the qubits on a square grid and routing them with SWAPs multiplies these counts by factors of 4 to more than 30.

Current square-grid and heavy-hex processors hold 100 to 156 qubits \cite{google2025surface,ibm_processor_types}, with two-qubit errors of order $10^{-3}$. At an error of $10^{-3}$ the expected number of faults reaches one after about $10^3$ gates, which we use only as a noise scale. A single all-to-all layer of any quartic instance exceeds this scale by a factor of 6 to $6\times10^{3}$, and QAOA-type or counterdiabatic circuits need several layers. Only the 3-XORSAT instance stays within the scale, with 400 gates per all-to-all layer, and on a grid even that does not. Higher-order experiments on 127- and 156-qubit heavy-hex processors accordingly used HUBOs whose three-body terms follow the qubit lattice, which avoids routing dense all-to-all interactions \cite{pelofske2024higher,romero2025bfdcqo_hubo,chandarana2025runtime}.

\paragraph{Trapped-ion processors.}
Ion registers are fully connected, so the routing factor disappears and the all-to-all gate counts of Table~\ref{tab:res_gate} apply. The largest register today holds 98 fully connected qubits with a two-qubit infidelity of $8\times10^{-4}$ \cite{ransford2025helios}, so the LABS instances and the 70-qubit portfolio fit into one register, the 100-qubit 3-XORSAT instance just exceeds it, and the bit-encoded budget instance does not fit. The same noise scale applies, and with two-qubit gates of order 200~$\mu$s executed largely serially, one layer takes seconds for LABS and 19~min for the portfolio, and would take hours for the budget instance.\looseness=-1

\paragraph{\Dirac{}.}
The native encoding, Eq.~\eqref{eq:poly}, uses 24 to 130 variables, with no user-supplied auxiliaries, penalties or compilation, and enforces the simplex constraint on the device itself. Under the mappings analyzed here, the embedding bounds for the portfolio and the budget allocation exceed the size of the cited annealers by more than an order of magnitude, and one layer of their phase-separation circuits exceeds the noise scale by more than three orders of magnitude. For the LABS instances the embedding bounds lie below the annealer sizes, but the quadratized coefficients span a range of several hundred, and one circuit layer exceeds the noise scale by a factor of 6 to 18. Of the four platforms considered, \Dirac{} is the one that accepts these six polynomials without reformulation. Acceptance does not imply resolution: the MVSK and budget polynomials exceed the stated analog resolution of the device (Sec.~\ref{sec:eqc}), a precision limit of the same kind as the one noted above for the annealer. How good the returned solutions are is the subject of Sec.~\ref{sec:results}.

\section{Results}
\label{sec:results}

\begin{table*}[t]
\caption{Results on \Dirac{} (schedule 3, 30 samples per instance), scored by independent re-evaluation against the reference of Table~\ref{tab:instances}, the certified optimum for 3-XORSAT and LABS and the best-known classical value for MVSK and budget. Gaps are absolute for 3-XORSAT (violated checks) and LABS (energy), relative for MVSK and budget. Device time is the per-job device time returned with the results, calibration excluded.}
\label{tab:results}
\footnotesize
\begin{ruledtabular}
\begin{tabular}{lrrrrrrrr}
Instance & Reference & Best & Median & Worst & Gap best & Gap median & s/sample & Device s \\
\hline
3-XORSAT $N{=}100$ & 0 & 3 & 10.5 & 14 & +3 & +10.5 & 2.09 & 63 \\
LABS $N{=}24$ & 36 & 56 & 74.0 & 104 & +20 & +38 & 0.27 & 8 \\
LABS $N{=}28$ & 50 & 66 & 96.0 & 114 & +16 & +46 & 0.51 & 15 \\
LABS $N{=}34$ & 65 & 129 & 157.0 & 201 & +64 & +92 & 1.00 & 30 \\
MVSK $N{=}35$ & $-6{,}867{,}311$ & $-6{,}845{,}117$ & $-6{,}650{,}416$ & $-6{,}435{,}015$ & +0.32\% & +3.16\% & 4.70 & 141 \\
Budget $N{=}130$ & $-4.3495$ & $-3.6882$ & $-3.2280$ & $-2.4988$ & +15.20\% & +25.78\% & 42.07 & 1{,}262 \\
\end{tabular}
\end{ruledtabular}
\end{table*}

\begin{figure*}[t]
\centering
\includegraphics[width=\textwidth]{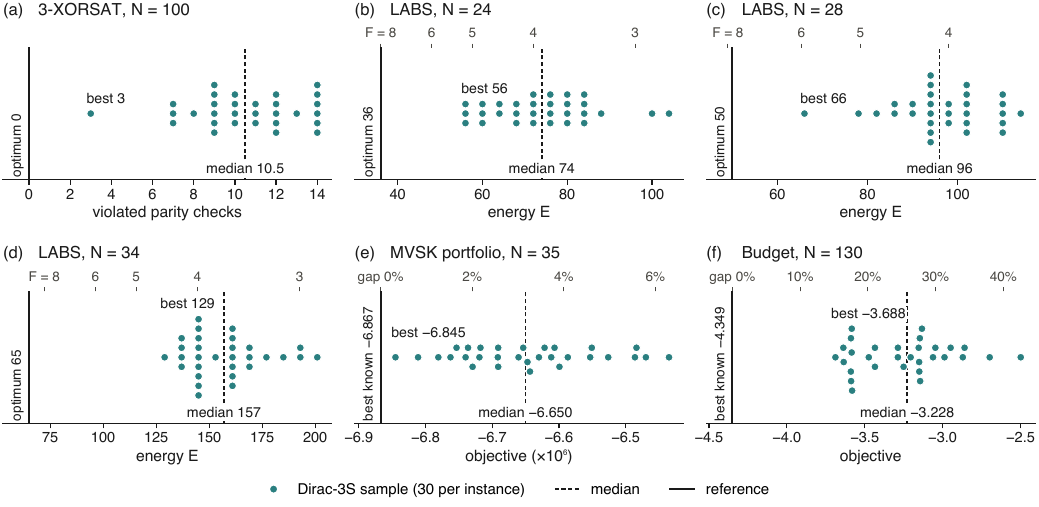}
\caption{Figure of merit of every returned sample (lower is better) on \Dirac{}, 30 samples per instance, for (a) 3-XORSAT, (b)--(d) LABS, (e) MVSK and (f) budget. The solid line is the reference, the dashed line the median, and the best sample is labeled. The upper axis gives the merit factor $F=N^{2}/2E$ for LABS and the relative gap to the best-known value for MVSK and budget.\looseness=-1}
\label{fig:samples}
\end{figure*}

\begin{figure*}[t]
\centering
\includegraphics[width=\textwidth]{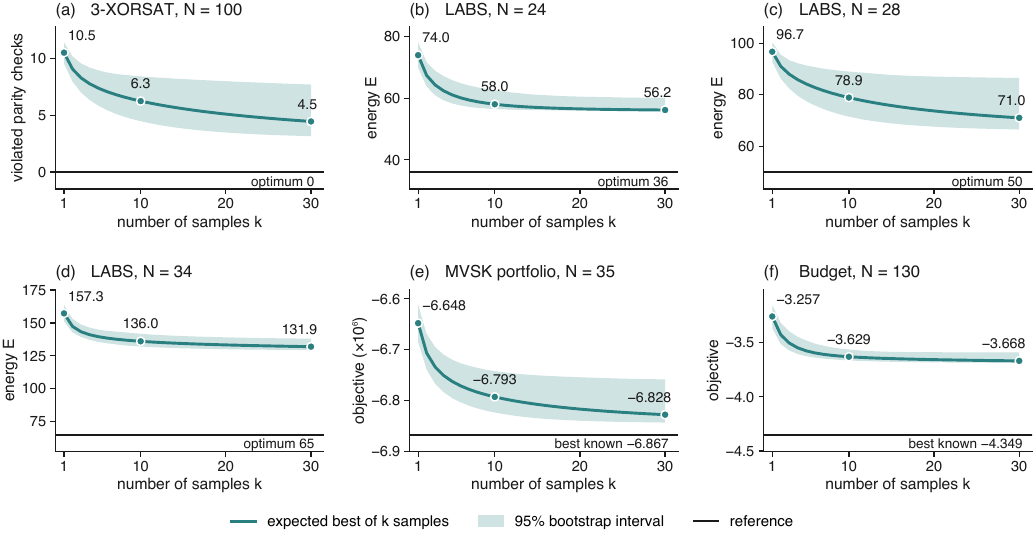}
\caption{Expected best-of-$k$ figure of merit $\Ek{k}$ (Appendix~\ref{app:methods}) under resampling from the empirical distribution of the 30 \Dirac{} samples, with 95\% bootstrap intervals, for (a) 3-XORSAT, (b)--(d) LABS $N{=}24$, 28 and 34, (e) MVSK and (f) budget. Dots mark $k=1$, $10$ and $30$. The curves summarize the observed samples and converge to the observed minimum by construction; they do not predict values below it.}
\label{fig:bestofk}
\end{figure*}

\begin{figure*}[t]
\centering
\includegraphics[width=\textwidth]{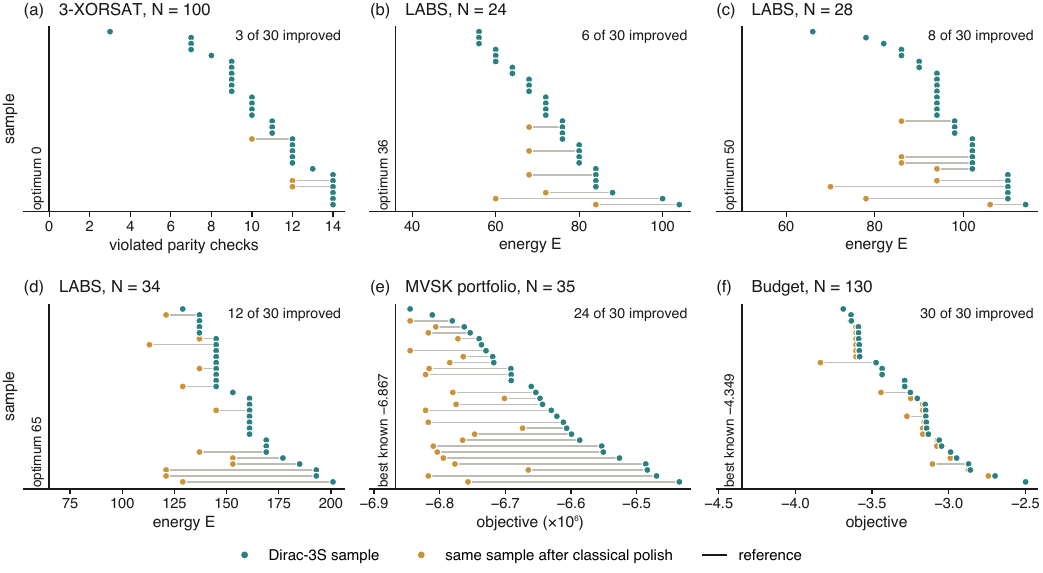}
\caption{Effect of a cheap classical polish started from each \Dirac{} sample, for (a) 3-XORSAT, (b)--(d) LABS $N{=}24$, 28 and 34, (e) MVSK and (f) budget. Samples are sorted by raw value and a connector links the raw and polished value of the same sample. Samples without a connector are local minima of the polish, except in (f), where every sample moves under SLSQP, most by less than the marker size. The polish is a best-improvement single- and double-flip descent for the binary instances, a single-lot and lot-transfer descent for MVSK, and SLSQP started from the sample for the budget instance.}
\label{fig:polish}
\end{figure*}

We ran all six instances natively on \Dirac{} with 30 samples each and re-scored every returned solution independently against its reference value (Appendix~\ref{app:methods}). Table~\ref{tab:results} and Fig.~\ref{fig:samples} give the best, median and worst sample of each instance, Fig.~\ref{fig:bestofk} the expected best of $k$ samples, Fig.~\ref{fig:polish} the effect of a cheap classical polish, and Table~\ref{tab:controls} the same polish applied to classical starting points. Every job completed, and every returned solution was distinct and feasible: integral and in range for the integer instances, and on the simplex to within $2\times10^{-7}$ for the budget instance. The results of this section come from a single job per instance, and the 30 samples of a job are treated below as exchangeable draws, an assumption a single job cannot test. Only the budget instance was run again later (Appendix~\ref{app:device}). The six jobs consumed 1{,}519~s of device time, 83\% of it for the continuous budget instance.\looseness=-1

On 3-XORSAT the best of 30 samples violates 3 of 100 checks and the median 10.5. All 30 samples are single-flip local minima, and a double-flip descent does not lower the best value. The parity-check matrix has full rank, so the planted assignment is the unique solution. The Hamming distances of the samples from it range from 35 to 59, close to those of uniformly random assignments (mean 50, standard deviation 5). The samples are therefore not partial decodings of the planted solution. Random assignments followed by the same descent reach similar medians, while the best device sample beats every classical batch minimum (Table~\ref{tab:controls}). The device thus behaves on this golf-course landscape as annealers and local heuristics do \cite{kowalsky2022xorsat}: it reaches configurations with a few violations, far from the unique solution. Since Gaussian elimination solves the instance directly (Sec.~\ref{sec:usecases}), it serves here as a probe of the samples, not as a computational task.\looseness=-1

For LABS the best energies are 56, 66 and 129 at $N{=}24$, 28 and 34, against optima of 36, 50 and 65, so the best samples reach 50\% to 76\% of the optimal merit factor. No two samples are related by the symmetries of the problem. Like the 3-XORSAT samples, all 90 LABS samples are single-flip local minima, so the returned configurations are locally optimized. Whether the photonic loop or a step of the processing pipeline produces this property cannot be read from the job record (Sec.~\ref{sec:eqc}). A double-flip descent improves a minority of the samples and never reaches the optimum, although at $N{=}34$ it lowers the best energy to 113 (Fig.~\ref{fig:polish}). In the matched control of Table~\ref{tab:controls}, the raw device medians lie below those of random sequences after a single-flip descent, and the polished device medians are close to those after the single- and double-flip descent. For the best of 30 the outcome is mixed: the polished device values lie above the median random-start batch at $N{=}24$ and 34, below it at $N{=}28$, and within the batch range in every case. One device batch against twenty classical ones therefore supports no statement of superiority in either direction.

\begin{figure*}[t]
\centering
\includegraphics[width=\textwidth]{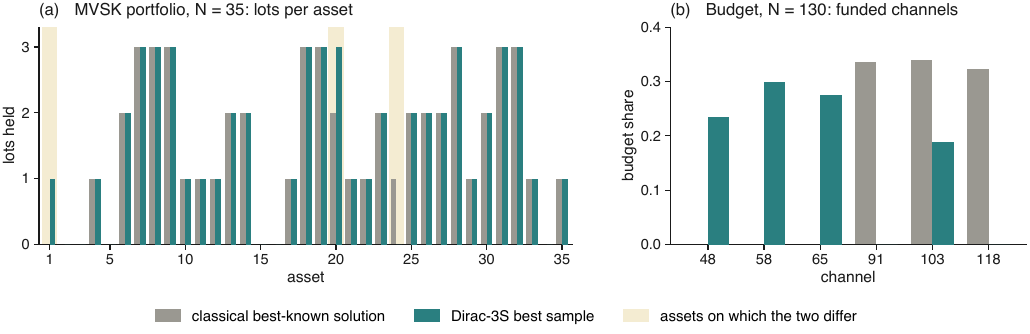}
\caption{Structure of the best \Dirac{} sample versus the classical best-known solution. (a) MVSK: lots per asset. The two solutions differ on three assets (shaded) and by one lot in total (54 versus 53). (b) Budget allocation: shares of the channels funded by either solution. The sample funds four channels and the best-known solution three, with channel 103 in common.}
\label{fig:structure}
\end{figure*}

The MVSK portfolio is the instance on which the device comes closest to the reference. The best sample lies within $0.32\%$ of the best-known value and the median within $3.2\%$ (Table~\ref{tab:results}); on the full, unsparsified objective the best sample lies within $0.39\%$. It differs from the best-known solution on three assets only (Fig.~\ref{fig:structure}a), and the samples use all four lot levels almost uniformly, without collapsing onto a binary subset.

Under single-lot and lot-transfer moves, 6 of the 30 samples are local minima and the descent improves the other 24. From random lot vectors the same descent ends at larger gaps than from the device samples (Table~\ref{tab:controls}). From no device sample, however, does it reach the best-known solution: the best configuration reached differs from that solution on three assets and is separated from it by a barrier. A move set chosen after this observation removes the barrier. A best-improvement descent over all changes of up to three assets by one lot each, up to 54{,}810 candidate moves per step, reaches the best-known solution from \MVSKREACH{} of the 30 samples in 1 to 8 steps. The same descent reaches it from 26 of 30 uniformly random lot vectors in 12 to 20 steps, and from the rounded minimizer of the convex relaxation in 3 steps (Table~\ref{tab:controls}). The device samples shorten this particular descent. Its end point, however, is reached about as often from random starts. Step counts are not a timing comparison, which would have to include the device time. It is the move set that finds the best-known solution.\looseness=-1

\begin{table*}[t]
\caption{Matched-start controls (Appendix~\ref{app:methods}): the cheap polish of Fig.~\ref{fig:polish} and the problem-specific descents of the text applied to the \Dirac{} samples and to classical starting points, uniformly random assignments or sign sequences (20 batches of 30), uniformly random lot vectors (30), the rounded minimizer of the convex relaxation (MVSK) and random SLSQP starts (600, of which 572 converged, in 20 batches of 30). Figure of merit as in Table~\ref{tab:results}: violated checks, LABS energy, relative gap to the best-known value. ``Median'' is the per-sample median after the polish; ``Best of 30'' is the best value of a batch of 30 (for the classical starts, the median over batches, with the range in brackets); the last column counts the starts whose polished value reaches the reference.\looseness=-1}
\label{tab:controls}
\footnotesize
\begin{ruledtabular}
\begin{tabular}{llllll}
Instance & Start & Polish & Median & Best of 30 & Reference reached \\
\hline
3-XORSAT $N{=}100$ & device & none & 10.5 & 3 & 0/30 \\
 & device & 1-flip, then 2-flip & 10 & 3 & 0/30 \\
 & random, 20 batches & 1-flip & 11 & 7 [5, 8] & 0/600 \\
 & random, 20 batches & 1-flip, then 2-flip & 11 & 7 [4, 8] & 0/600 \\
LABS $N{=}24$ & device & none & 74 & 56 & 0/30 \\
 & device & 1-flip, then 2-flip & 70 & 56 & 0/30 \\
 & random, 20 batches & 1-flip & 84 & 58 [52, 68] & 0/600 \\
 & random, 20 batches & 1-flip, then 2-flip & 72 & 52 [36, 60] & 2/600 \\
LABS $N{=}28$ & device & none & 96 & 66 & 0/30 \\
 & device & 1-flip, then 2-flip & 94 & 66 & 0/30 \\
 & random, 20 batches & 1-flip & 110 & 78 [50, 90] & 1/600 \\
 & random, 20 batches & 1-flip, then 2-flip & 98 & 74 [50, 90] & 2/600 \\
LABS $N{=}34$ & device & none & 157 & 129 & 0/30 \\
 & device & 1-flip, then 2-flip & 145 & 113 & 0/30 \\
 & random, 20 batches & 1-flip & 169 & 117 [89, 137] & 0/600 \\
 & random, 20 batches & 1-flip, then 2-flip & 145 & 109 [89, 129] & 0/600 \\
MVSK $N{=}35$ & device & none & 3.16\% & 0.32\% & 0/30 \\
 & device & lot / transfer & 1.24\% & 0.32\% & 0/30 \\
 & device & three-asset & 0.00\% & 0.00\% & 27/30 \\
 & random, 30 & lot / transfer & 1.66\% & 0.57\% & 0/30 \\
 & random, 30 & three-asset & 0.00\% & 0.00\% & 26/30 \\
 & rounded relaxation & three-asset & -- & 0.00\% & 1/1 \\
Budget $N{=}130$ & device & none & 25.78\% & 15.20\% & 0/30 \\
 & device & SLSQP & 24.53\% & 11.78\% & 0/30 \\
 & random, 20 batches & SLSQP & 18.96\% & 7.71\% [4.31, 10.50] & 0/572 \\
\end{tabular}
\end{ruledtabular}
\end{table*}

On the budget instance, the one continuous problem, the best sample lies $15.2\%$ above the best-known value and the median $25.8\%$ (Table~\ref{tab:results}). Every sample is on the simplex and sparse, funding 4 to 10 of the 130 channels with a share above $10^{-6}$, and the best sample shares one of its four channels with the three-channel best-known solution (Fig.~\ref{fig:structure}b). Every sample moves under SLSQP, so none of the continuous samples is a stationary point of the constrained problem, and the best polished sample still lies $11.8\%$ above the best-known value. A local search over sparse supports, which adds, drops or swaps one channel and re-solves the restricted problem with SLSQP, converged from the 15 best samples to the same local optima as the SLSQP polish. Random SLSQP starts also miss the best-known basin but do better than the device-seeded polish: in 20 batches of 30 starts the batch best lies below the polished device best every time (Table~\ref{tab:controls}). On this instance the device supplies feasible sparse allocations but no better starting points than random ones. Three further runs of this instance, one of them at schedule~2, returned sample values consistent with these, and again no sample reached the reference (Appendix~\ref{app:device}).

The best-of-$k$ curves of Fig.~\ref{fig:bestofk} describe the 30 observed values only. With no sample at the reference, the per-sample probability of reaching it is not shown to be zero: under independent trials it is bounded above by 9.5\% at 95\% confidence. Device time per sample grows with problem size but is not set by the term count alone. Among the quartic integer-mode instances it rises from 0.27~s for LABS $N{=}24$ to 4.7~s for the portfolio, the sparse 3-XORSAT instance takes 2.1~s, and the continuous budget instance is an order of magnitude slower still (Table~\ref{tab:results}).\looseness=-1

\section{Discussion and conclusion}
\label{sec:discussion}

In this work we have formulated four use cases whose objectives are higher-order by nature as six cubic and quartic instances over binary, integer and simplex-constrained continuous variables and run them natively on \Dirac{}. Every returned solution was scored against a certified optimum or an independently established best-known value, and the cost of the same instances was estimated for quadratic annealers, superconducting processors and trapped-ion processors. Under the mappings analyzed, the portfolio and the budget instance have embedding bounds far above the size of the cited annealers and circuit layers orders of magnitude above the noise scale. For the LABS instances the embedding bounds lie below the annealer sizes, but the quadratized coefficients span a range of several hundred (Sec.~\ref{sec:resources}).

Every sample is feasible and distinct, and every binary sample is a single-flip local minimum. The quality of the binary samples is comparable to that of locally optimized random starts (Table~\ref{tab:controls}). With one instance per class no trend can be established, but the quality of the configurations is consistent with the structure of each landscape. On the portfolio, whose unrounded objective is convex (Sec.~\ref{sec:usecases}), the best sample lies within 0.32\% of the best-known value despite the largest term count and dynamic range of the integer-mode instances. On LABS the samples lie in the rugged region of locally optimal codes that also slows classical heuristics, and on the golf-course landscape of planted 3-XORSAT they lie as far from the unique solution as random assignments. On the continuous budget instance the device finds sparse, feasible allocations but not the basin of the best-known solution, which random classical restarts also miss. The weakest results are on the largest quartic binary problem, LABS $N{=}34$, and on the continuous problem, which also has the longest device time per sample. They mark where improvements to the device would matter most.

Classical local search improves on the device samples: a cheap polish lowered the best LABS $N{=}34$ and budget values, and a move set adapted to the problem carried \MVSKREACH{} of the 30 MVSK samples to the best-known solution. Whether a device sample is a better start than a random one depends on the problem (Table~\ref{tab:controls}). Device starts shorten the MVSK descent but reach the reference about as often as random starts, the LABS best of 30 is better than the control at one size and worse at two, random starts polish to better budget values, and on 3-XORSAT no tested descent reaches a satisfying assignment from either start. An improvement in final quality or in total solution time, which would have to include device time and preprocessing, is not established.

Four practical points follow for a user of such a device. A cubic or quartic objective can be submitted as it is, but its coefficients should be scaled and pruned with the 200:1 resolution in mind; the portfolio shows how much weight a long tail of small terms can carry (Sec.~\ref{sec:eqc}). Binary and integer samples arrive as single-flip local minima, so a generic polish adds little, while a move set matched to the problem adds much more. The best value still improves at 30 samples on some instances and flattens after about ten on others (Fig.~\ref{fig:bestofk}), so the sample count is worth tuning per problem. Finally, on the one instance where two schedules were compared, the faster schedule did no worse (Appendix~\ref{app:device}).

At the sizes studied here classical solvers reach the reference values in seconds to minutes (Appendix~\ref{app:methods}), in line with an earlier independent assessment of \Series{} \cite{moosavian2026critical}. No quantum advantage is claimed, nor any contribution of a nonclassical resource (Sec.~\ref{sec:eqc}). The advantage established here is one of expressibility and mapping cost relative to quadratic annealers and to cost-Hamiltonian circuits on superconducting and trapped-ion processors, under the mappings analyzed. It is a property of the input format, shared with classical higher-order solvers and with other higher-order Ising machines \cite{bybee2023higher,ahsan2026neuromorphic}, and not a performance result.

The limitations of this study concern the instances, the device runs, the reference values and the resource estimates. Each use case is represented by one instance (three sizes for LABS), and the portfolio and budget data are synthetic. Each instance was run once with 30 samples at schedule~3; only the budget instance was repeated, with consistent results (Appendix~\ref{app:device}). Run-to-run variability of the other instances and the quality--time trade-off across schedules are therefore not characterized, and the bootstrap intervals of Fig.~\ref{fig:bestofk} describe within-job resampling only. The photon-number and fluctuation parameters were left at the presets of the schedule, which are not stored in the job record, and the correction step applied to the returned samples is not specified (Sec.~\ref{sec:eqc}).

The MVSK and budget references are best-known values rather than certified optima. On the full unrounded objective the MVSK reference is certified to lie within 2.7\% of the integer optimum, with no such certificate for the rounded, sparsified polynomial. The budget reference is certified only as the best three-channel allocation, and random multistart did not find it (Appendix~\ref{app:instances}).

The resource estimates rest on simple models with stated assumptions and specific mappings. The embedding bounds hold for the quadratizations used here, not for every encoding, and the gate counts are construction costs that compilation can reduce (Appendix~\ref{app:resources}).

With the expanded envelope of \Dirac{} (Fig.~\ref{fig:envelope}), the measurements that matter most can be made on the same device. A scaling sweep in $N$ on LABS and on the budget problem, with instance ensembles and fixed quality targets, would compare the growth of device time at fixed quality with that of classical time to solution, which is exponential for the LABS and 3-XORSAT heuristics \cite{kowalsky2022xorsat,cadavid2025qemts}. A sweep of the schedule and of the photon-number and fluctuation parameters at fixed instance, with repeated jobs, would supply the statistics that a single job cannot. Quintic instances, together with a same-device comparison of a native and a quadratized formulation of one instance, would isolate the cost of quadratization. A controlled test of device samples as starts for a classical memetic or tabu search \cite{cadavid2025qemts}, against random and heuristic starts at equal total budget, would settle the value of hybrid use, for which the matched controls of Table~\ref{tab:controls} give no consistent indication at the present sizes. Together these experiments would show whether the advantage in expressibility extends to time to solution at sizes where classical solvers slow down.

\section*{Data availability}
The data and code that support the findings of this study are available from the corresponding authors upon reasonable request.

\appendix

\section{Methods}
\label{app:methods}

Beyond the relaxation schedule (3) and the sample count (30 per instance), all solver parameters (mean photon number, quantum-fluctuation coefficient, solution precision) were left unset, so the device applied the presets of the schedule, which are not stored in the job record, and no warm start, problem-specific tuning or post-selection was used. Appendix~\ref{app:device} lists modes, levels, job parameters and timings.\looseness=-1

The device reports an energy for every sample, but none of the numbers in the main text rely on it. Every returned solution was re-evaluated in double precision from the polynomial file with the constant added back, and the figure of merit was computed from that value: violated checks $E/2$ for 3-XORSAT, energy $E$ for LABS, with the merit factor $F$ quoted alongside, and the objective itself for MVSK and budget. Integer solutions were checked to be integral and in range, continuous solutions to lie on the simplex. The device's own estimate agreed with the re-evaluated value to $8\times10^{-3}$ for the binary instances, to $5.5$ for MVSK (below $10^{-6}$ relative) and to $1.1\times10^{-6}$ for the budget instance.

Because each sample is a separate run of the relaxation loop and no solution was returned twice, we summarize the 30 values also by the expected best of $k$,
$\Ek{k} = \sum_{i=1}^{n} E_{(i)} \big[(\tfrac{n-i+1}{n})^{k} - (\tfrac{n-i}{n})^{k}\big]$,
computed exactly from the sorted sample values $E_{(1)} \le \dots \le E_{(n)}$ of the figure of merit under i.i.d.\ resampling from the empirical distribution. This is a description of the observed sample, conditional on it: $\Ek{k} \to E_{(1)}$ as $k$ grows, so the curves of Fig.~\ref{fig:bestofk} saturate at the observed minimum by construction and cannot indicate values below it. The shaded bands of Fig.~\ref{fig:bestofk} are 95\% percentile intervals from 4{,}000 bootstrap resamples of the 30 samples; they describe within-job resampling only; job-to-job variation cannot be estimated from a single job. Gaps to the reference are absolute differences for 3-XORSAT and LABS and relative differences for MVSK and budget.

A sample that is not a local minimum can be improved at negligible cost, so for each returned solution we ask whether a cheap classical move set can improve it and, if so, where a best-improvement descent from it ends. The move sets are single flips, followed by double flips once no single flip improves, for the binary instances; changes of one asset to any other lot level and lot transfers $x_i{+}1, x_j{-}1$ for MVSK (the single-lot and lot-transfer moves of the text); and SLSQP started from the sample for budget. Figure~\ref{fig:polish} shows the effect on every sample.

To calibrate the polish, the same descents were applied to the classical starting points listed in Table~\ref{tab:controls}, all drawn with fixed seeds; the random SLSQP starts are the 600 of Appendix~\ref{app:instances}. The upper bound of 9.5\% on an unobserved per-sample success probability, $1 - 0.05^{1/30}$, assumes independent and identically distributed trials.

The reference values against which the samples are scored come from classical computation: the planted assignment for 3-XORSAT, exhaustive branch-and-bound for LABS \cite{mertens1996exhaustive,packebusch2016labs}, problem-tuned simulated annealing \cite{kirkpatrick1983sa} for MVSK, where an independent integer parallel-tempering run \cite{hukushima1996pt} on the unpruned polynomial came within 0.6\% and found nothing lower, and seeded SLSQP \cite{kraft1988slsqp,virtanen2020scipy} confirmed by an exhaustive support search for budget (Sec.~\ref{sec:usecases}, Appendix~\ref{app:instances}). The LABS optima are taken from the literature; the MVSK and budget references take seconds to minutes to compute on a laptop.

\section{Job parameters and timings}
\label{app:device}
The jobs of 28~August were submitted with \texttt{eqc-models} 0.21.0 \cite{qci_eqc_models}. Integer-mode jobs (3-XORSAT, LABS, MVSK) were run with \texttt{num\_levels} equal to 2 for every binary variable and 4 for every MVSK variable, \texttt{relaxation\_schedule} $=3$ and \texttt{num\_samples} $=30$. The budget job was run in continuous mode with \texttt{sum\_constraint} $=1.0$, \texttt{relaxation\_schedule} $=3$ and \texttt{num\_samples} $=30$. Reported calibration times were 3.3~s (3-XORSAT), 0~s (LABS), 10.0~s (MVSK) and 10.1~s (budget). Device and wall times per job were 62.7/105~s, 8.1/31~s, 15.4/61~s, 30.1/121~s, 141.0/761~s and 1{,}262/1{,}281~s for 3-XORSAT, LABS $N{=}24, 28, 34$, MVSK and budget. Each sample record contains the device energy, float32 and float64 host evaluations, the figure of merit, feasibility flags and the solution vector. The float64 host value agreed exactly with our independent re-evaluation for every integer-mode sample and to $2\times10^{-15}$ for the continuous samples.

\begin{figure}[t]
\centering
\includegraphics[width=\columnwidth]{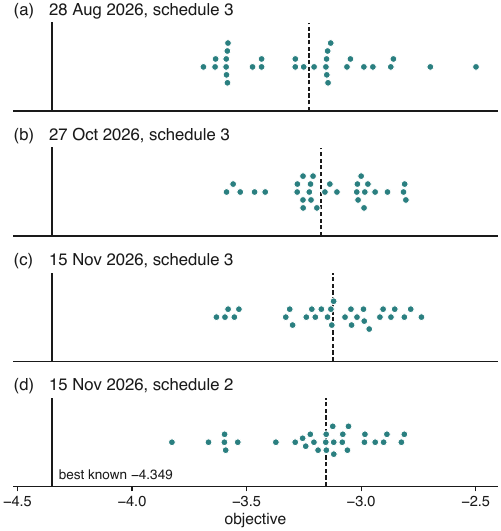}
\caption{Repeated runs of the budget instance on \Dirac{}, 30 samples each: (a) the main-text run, (b) and (c) two more runs at schedule~3, and (d) a run at schedule~2. The solid line is the best-known value and the dashed line the median of each run. Panels (c) and (d) show device-reported values.}
\label{fig:repeat}
\end{figure}

\paragraph{Repeated runs of the budget instance.} The budget instance was run three more times with 30 samples each: on 27~October and 15~November~2026 at schedule~3, and on 15~November at schedule~2 (Fig.~\ref{fig:repeat}). The four runs support the same conclusions. The best samples lie 12.0\% to 17.5\% above the best-known value and the medians 25.8\% to 28.2\%; pooled over all 120 samples, the best lies 12.0\% and the median 27.5\% above it. The run of the main text has the lowest median of the four. A Kruskal--Wallis test over the four runs gives $p = 0.26$, and two-sided Mann--Whitney tests of each later run against the run of the main text give $p$ between 0.07 and 0.40. The schedule-2 run took 31~s per sample, against 42 to 46~s at schedule~3, and returned the best single sample, $-3.826$. No sample of any run reached the reference, which lowers the bound on the per-sample probability of reaching it from 9.5\% to 2.5\% for this instance. The November values are the energies returned with the jobs; host re-evaluation of the best sample of each November run and of all samples of the two earlier runs agreed with the returned energies to within $2\times10^{-6}$.

\section{Instance construction}
\label{app:instances}
Every exported polynomial was checked against a direct implementation of its objective on the reference solution and on random points. The instance files were frozen before the run on \Dirac{} and are identified by SHA-256 hashes recorded in the device output.
\paragraph{3-XORSAT.} Configuration-model 3-regular hypergraph on 100 variables with 100 checks, resampled until no check contains a repeated variable. The planted assignment is drawn uniformly and the parities $\sigma_m$ are set to satisfy it (seed 1). Coefficients are integers in $[-8, 8]$.
\paragraph{LABS.} Exact multilinear expansion of Eq.~\eqref{eq:labs} with $s_i = 1 - 2x_i$. The exported polynomial reproduces $E$ on the optimum and on 5{,}000 random sequences. The dropped constants are 4{,}324, 6{,}930 and 12{,}529.
\paragraph{MVSK.} Returns: 600 observations, four Student-$t_5$ factors with Gaussian loadings, Student-$t_6$ idiosyncratic noise and a zero-mean $\chi^2$-type shock for asymmetry, standardized and winsorized at $\pm6$. Reward $\mu_i \sim U(0.5, 3)$. The polynomial is assembled from the central co-moment tensors with multinomial multiplicities so that it equals the tensor objective exactly (relative error below $10^{-9}$), then scaled by $10^4$ and rounded. Brute force at $N{=}5, 6, 7$ confirms that rounding preserves the minimizer at those sizes. The sparsification threshold is $10^{-3}$ of the largest coefficient and removes 3{,}080 of 82{,}244 terms. At $N{=}35$ preservation of the minimizer under rounding and sparsification is not certified; the best-known solution is the same for the sparse and the full polynomial, whose values at that point differ by 15{,}658 (0.23\%), and the rounding changes the value by 1{,}009 (0.015\%). Because the polynomial contains integer powers ($x_i^2, x_i^3, x_i^4$), the term counts in Table~\ref{tab:instances} exceed the number of index subsets.

\emph{Convex relaxation.} With the $1/T$-normalized central co-moments the unrounded objective is $E(x) = -\lambda_1\mu^{\!\top}x + T^{-1}\sum_t \varphi(r_t^{\top}x)$ with $r_t$ the centered return vector of observation $t$ and $\varphi(z) = z^2 - 0.3z^3 + 0.1z^4$, whose second derivative $1.2(z-0.75)^2 + 1.325$ is positive, so $E$ is convex on $\mathbb{R}^{35}$ and can be evaluated in $O(TN)$ operations. L-BFGS-B with the analytic gradient converges from three starting points to a point $y$ at which the scaled objective $f$ takes the value $f(y) = -7{,}069{,}403.8$. Since $f$ is convex, the tangent-plane bound $L = f(y) + \sum_i \min\{-g_i y_i,\, g_i(3-y_i)\}$ with $g = \nabla f(y)$ is a certified lower bound on $\min_{x\in[0,3]^{35}} f(x)$; it evaluates to $L = -7{,}069{,}404.0$, within 0.13 of $f(y)$. The best-known solution evaluates to $-6{,}881{,}960$ on the unrounded objective, 2.7\% above $L$, so the integer optimum of the unrounded objective lies within 2.7\% of it. No bound is transferred to the rounded, sparsified polynomial: that would need a uniform bound on the pruned part over the box, which the difference of 15{,}658 at a single point does not provide. The relaxation minimizer holds 54.0 lots with 5 assets at 0 and 4 at 3; rounded to the nearest lots it holds 53 lots in 27 assets, differs from the best-known solution on 4 assets, scores $-6{,}240{,}629$ on the sparse polynomial, and reaches $-6{,}867{,}311$ after 3 steps of the three-asset descent. The outcomes of the same descents from random lot vectors are in Table~\ref{tab:controls}.\looseness=-1
\paragraph{Budget.} $a_i \sim U(0.5, 2)$, $b_i \sim U(0.2, 1)$, $g_{ijk} \sim 16\,\mathcal{N}(0,1)$ (seed 1), with synergies below $10^{-3}$ of the largest coefficient zeroed. Non-convexity on the simplex was verified through the Hessian projected onto the tangent space $\{d: \sum_i d_i = 0\}$, which has a negative eigenvalue (between $-47$ and $-38$) at all of 50 random simplex points. Of 600 random SLSQP starts (Dirichlet-distributed initial points, tolerance $10^{-11}$), 572 converged with a projected-gradient residual below $7\times10^{-5}$; they gave 162 distinct objective values at a clustering tolerance of $10^{-5}$ (161 at $10^{-4}$, 136 at $10^{-3}$); 8 starts ended at $-4.162045$ and none at the best-known $-4.349464$. At every converged end point the Hessian projected onto the tangent space of the active face is positive definite (smallest eigenvalue at least 0.94) and the multiplier condition holds for the zero coordinates, so all 572 are strict local minima of the simplex problem. Grouped in order into 20 batches of 30 starts, the batch minima have median $-4.014$ and range $-4.162$ to $-3.893$.

The best-known value was found by SLSQP from the barycenters of the 1{,}500 highest-synergy triples (5 of 1{,}500 seeds). It is certified as the minimum over all three-channel supports: on a support $\{i,j,k\}$ the objective restricted to the simplex is a cubic in two variables whose only cubic term is $-g_{ijk}v_iv_jv_k$. Its interior stationary points are the common roots of two quadratics, found by eliminating one variable (a quartic, solved as a companion-matrix eigenproblem and polished by Newton steps to residuals below $10^{-13}$); on the three edges the cubic term vanishes and the quadratic minima are explicit; and the vertices are evaluated. The minimum over these candidates for all 357{,}760 supports is $-4.349464$ on $\{91, 103, 118\}$. As an independent check, a multistart SLSQP run on every support matched the certified minimum on all but 103 supports, where it stopped higher by up to 0.045, and was never below it. A dense grid on the 200 best supports found nothing below the certified values. The greedy four- and five-channel extensions of the 300 best supports, solved by the same multistart SLSQP, gave nothing lower, without a certificate.

\section{Resource-estimation models}
\label{app:resources}
\paragraph{Encoding.} Binary variables are used as they are. An integer $x_i \in \{0,\dots,3\}$ is written $x_i = b_{i0} + 2b_{i1}$ and every monomial is expanded into multilinear bit-monomials ($b^2 = b$), merging equal monomials. A continuous share is written as a $B$-bit fixed-point number $v_i = \sum_{a=1}^{B} 2^{-a} b_{ia}$, whose sum over channels equals 1 only on a subset of bit patterns that the simplex penalty $(\sum_i v_i - 1)^2$ selects; the penalty contributes a coupler for every pair of bits. The cubic bit-monomials, $B^3$ per cubic term, involve three different channels and are all distinct; the fixed-point weights spread the coefficients of each cubic term over a factor $2^{3(B-1)}$, 512 at 4 bits and $2\times10^6$ at 8 bits.
\paragraph{Quadratization.} Each monomial of degree 3 is written $y_{ab} x_c$ and each of degree 4 as $y_{ab} y_{cd}$ with auxiliaries $y_{ab} = x_a x_b$. The pairs are chosen by a greedy cover that prefers pairs already in use (three refinement passes), so the auxiliary count is an upper bound on the minimum. Each auxiliary carries the Rosenberg penalty $x_ax_b - 2x_ay - 2x_by + 3y$ with a weight equal to the sum of the absolute coefficients of the monomials routed through $y$ plus the smallest coefficient, a sufficient condition for the penalty to dominate any gain from violating $y = x_ax_b$. Couplers are the distinct variable pairs of the resulting QUBO, including the three penalty couplers per auxiliary. Penalty-weight amplification is the ratio of the largest penalty weight $P$ to the largest coefficient of the binary-encoded polynomial before quadratization. The gadget then contains couplers of weight $-2P$ and a linear term $3P$, so the largest coupler of the final QUBO is twice and its largest linear term three times the quoted factor; the coefficient range quoted in Sec.~\ref{sec:resources} is the ratio of the largest final coefficient to the smallest coefficient before quadratization. For the portfolio the bit expansion alone widens this range from 997 to $9.5\times10^5$, and the penalties contribute the remaining factor of 465. For the budget instance the channels are split into two groups of 65, every channel triple contains a same-group pair, and one auxiliary per within-group bit pair, $2\binom{65}{2}B^2 = 4{,}160\,B^2$ in all, covers every cubic bit-monomial; each monomial then contributes one coupler between its auxiliary and its third bit. Table~\ref{tab:res_annealer} lists the exact counts of this construction with all auxiliaries allocated, $L = 130B + 4{,}160\,B^2$ logical variables and $C = 356{,}310\,B^3 + 2A + 65B(130B-1)$ couplers with $A$ the number of auxiliaries; they are construction counts, not bounds on the minimum quadratization.
\paragraph{Embedding.} A chain of $m$ physical qubits \cite{choi2008minor} on a hardware graph of vertex degree $\Delta$ \cite{boothby2020pegasus,boothby2021zephyr} is connected, so it spends at least $m-1$ couplers on itself and offers at most $\Delta m - 2(m-1) = (\Delta-2)m + 2$ external coupler endpoints. A logical node of degree $\delta$ therefore occupies at least $\max\{1, \lceil(\delta-2)/(\Delta-2)\rceil\}$ physical qubits, $\lceil(\delta-2)/13\rceil$ at degree 15 and $\lceil(\delta-2)/18\rceil$ at degree 20. Summing over logical nodes gives the lower bounds quoted for degree 15; for the budget instance, whose node degrees we did not enumerate, we use the aggregate bound $\max\{L, \lceil(2C-2L)/(\Delta-2)\rceil\}$ with $L$ logical nodes and $C$ couplers. At degree 20 the bounds are 555, 844 and 1{,}426 for LABS, 106{,}897 for MVSK and $2.6\times10^6$ and $2.0\times10^7$ for the budget instance at 4 and 8 bits; the MVSK and budget bounds are far above the roughly 4{,}400 qubits of the largest degree-20 annealer \cite{dwave2025adv2}, whereas a bound below the capacity, as for LABS, does not by itself show that an embedding exists.\looseness=-1
\paragraph{Gate counts.} The cost Hamiltonian is converted to the Pauli basis by $x_i = (1-Z_i)/2$: a Boolean monomial of degree $d$ generates the $2^d$ $Z$-strings on the subsets of its variables, identical strings are merged and vanishing ones dropped. The identity between the Boolean polynomial and its $Z$-string expansion was checked numerically on 100 random configurations for each binary instance (relative error below $10^{-9}$). For LABS the expansion reproduces the spin form of Eq.~\eqref{eq:labs}: the odd strings cancel and the quadratic strings $s_is_{i+2k}$ number $\sum_k (N-2k) = 132$, 182 and 272 at $N{=}24$, 28 and 34, next to 946, 1{,}547 and 2{,}856 quartic strings; the 3-XORSAT Hamiltonian is its 100 three-body strings. For the bit-encoded portfolio the expansion of the actual rounded, pruned polynomial has 948{,}996 nonzero strings (70, 2{,}415, 54{,}740 and 891{,}771 of weights 1 to 4) against 948{,}995 bit-monomials, one bit-monomial canceling while its string survives through its supersets; all coefficients are integer multiples of $1/16$, so the zero test is exact. For the budget encodings the cubic strings correspond exactly to the $356{,}310\,B^3$ cubic bit-monomials, and the lower-weight strings number at most $130B + 65B(130B-1)$, below 0.6\% of the total.

A string of weight $w$ contributes $2(w-1)$ two-qubit gates per layer. On a square grid of $\ell = \lceil\sqrt{Q}\rceil$ qubits a side, with $Q$ the number of qubits, the $w-1$ partner qubits of a string are moved next to a target by SWAP gates (three two-qubit gates each) over the mean Manhattan distance $\bar D = 2(\ell^2-1)/3\ell$, counted one way, which gives $(w-1)\,[2 + 3(\bar D - 1)]$ two-qubit gates per string. Heavy-hex lattices have longer paths by roughly a factor 1.5.

Rotations whose parities share a prefix can share CNOTs ($Z_1Z_2$ and $Z_1Z_2Z_3$ take four CNOTs together instead of six); native two-qubit rotations change the accounting, and with one $ZZ$ rotation and four CNOTs per quartic string, as in the LABS experiment of Ref.~\cite{shaydulin2024labs}, the LABS layers take 4{,}862, 7{,}917 and 14{,}552 two-qubit gates before any cancellation; generalized swap networks \cite{ogorman2019swap} remove the size-dependent routing factor of the grid estimate; and compiled objective-evaluation circuits \cite{bako2025progqaoa} can exploit further structure. We make no claim about optimally compiled circuit sizes; the gaps of Table~\ref{tab:res_gate} are those of the stated construction. Layer times assume serial execution at 200~$\mu$s per two-qubit gate for trapped ions, of the order of the gate duration itself (ion transport, which sets the layer time of current shuttling registers \cite{ransford2025helios}, is not included), and 0.3~$\mu$s for superconducting processors.\looseness=-1

\clearpage
\bibliography{refs}

\end{document}